\documentclass[times,authoryear]{elsarticle}
\usepackage{jasr}
\usepackage{framed,multirow}
\usepackage{amssymb}
\usepackage{latexsym}
\usepackage[switch]{lineno}
\usepackage{url}
\usepackage{ulem}
\usepackage{xcolor}
\definecolor{newcolor}{rgb}{.8,.349,.1}
\usepackage[citebordercolor=white]{hyperref}
\def\oiii{[{O}{\sc iii}]$\lambda\-\lambda$\-4959,\-5007\/}

\def\civ{{C}{\sc iv}$\lambda$1549\/}
\def\mgii{{Mg}{\sc ii}$\lambda$2800\/}

\def\hb{H$\beta$\/}

\def\heiiuv{{He}{\sc ii}$\lambda$1640\/}
\def\feii{{Fe}{\sc ii}\/}
\def\kms{km\,s$^{-1}$\/}

\def\ciii{{C}{\sc iii}]$\lambda$1909\/}

\def\rfe{$R_\mathrm{{Fe}{\textsc{ii}}}$\/}
\def\mbh{$M_\mathrm{\rm BH}$\/}
\def\lledd{$L/L_\mathrm{\rm Edd}$\/}

\def\c14{$c$(1/4)\/}
\def\ergs{erg s$^{-1}$\/}
\def\apjs{ApJS}
\def\aap{A\&Ap}
\def\mnras{MNRAS\/}
\def\apjl{ApJL\/}
\def\apj{ApJ\/}

\def\aj{AJ\/}
\def\nat{Nature\/}
\def\apss{ApSS\/}

\def\araa{ARA\&Ap}
\def\nar{NARev}
\def\prd{PhysRevD}
\def\prl{PhysRevLett}

\defcitealias{terefemengistueetal23}{M23}
\defcitealias{terefemengistueetal24}{M24}
\journal{Advances in Space Research}

\begin{document}

\verso{Given-name Surname \textit{etal}}

\begin{frontmatter}

\title{The Disk Plus (Failed) Wind System of 3C 47:\\ A Story of Accretion Disks and Binary Black Holes\tnoteref{tnote1}}%
\tnotetext[tnote1]{Mainly based on the invited talk presented at the VI Conference on AGN and GLs.}

\author[1]{P.   \snm{Marziani}\corref{cor1}}
\cortext[cor1]{Corresponding author: 
  Tel.: +39-049-829-3415;  
  fax: +39-049-875-9840;}
\author[2,3]{S. \snm{Terefe Mengistue}
}
\author[4]{A. \snm{del Olmo}}
\ead{chony@iaa}
\author[4,3]{M.   \snm{Povi\'c}}
\ead{mpovic@iaa.es}
\author[4]{J. \snm{Perea}}
\ead{jaime@iaa.es}
\author[5]{S. \snm{Komossa}}
\ead{skomossa@mpifr.de}
\author[6]{E. \snm{Bon}}
\ead{ebon@aob.rs}
\author[6]{N.  \snm{Bon}}
\ead{nbon@aob.rs}
\author[6]{L. \v C \snm{Popovi\'c}}
\author[4]{A. \snm{Deconto-Machado}}
\ead{alice.deconto@inaf.it}
\author[4]{I.   \snm{Marquez}}
\ead{isabel@iaa.csic.es}
\author[7]{M. A.  \snm{Mart\'{\i}nez Carballo }}
\ead{gelimc@unizar.es}

\affiliation[1]{organization={INAF - Osservatorio Astronomico di Padova},
                addressline={Vicolo Osservatorio 5},
                city={Padova},
                postcode={IT35122},
                country={Italy}}

\affiliation[2]{organization={Space Science and   Geospatial  Institute (SSGI)},
                addressline={PO Box 33679},
                city={Addis Ababa},
                country={Ethiopia}}

\affiliation[3]{organization={Physics Department, College of Natural Sciences, Jimma University, PO Box 378, Jimma, Ethiopia}}

\affiliation[4]{organization={Instituto de Astrofísica de Andalucía (CSIC)},
                addressline={Glorieta de Astronom\'\i a s/n},
                city={Granada},
                postcode={E18008},
                country={Spain}}

\affiliation[5]{organization={Max-Panck Institut f\"ur Radioastronomie},
                addressline={Auf dem H\"ugel 69},
                city={Bonn},
                postcode={D53121},
                country={Germany}}

\affiliation[6]{organization={Astronomical Observatory of Belgrade},
                addressline={Volgina 7},
                city={Belgrade},
                postcode={11000},
                country={Serbia}}

\affiliation[7]{organization={IUMA, CoDy and Dpto. Matemática Aplicada, Universidad de Zaragoza},
                city={Zaragoza},
                postcode={E-50009},
                country={Spain}}

\received{\ldots}
\finalform{\ldots}
\accepted{\ldots}
\availableonline{\ldots}
\communicated{S. Sarkar}

\begin{abstract}
Optically thick, geometrically thin accretion disks around supermassive black holes are thought to contribute to broad-line emission in type-1 active galactic nuclei (AGN). However, observed emission line profiles most often deviate from those expected from a rotating disk, and the role of accretion disks in contributing to broad Balmer lines and high-ionization UV lines such as CIV$\lambda$1549 in radio-loud (relativistically ``jetted'') AGN remains unclear. {  This report builds on the findings of three previous studies and places them within a broader context, offering new insights and a more coherent interpretation of earlier   results and their implications.} {  It examines} the role of accretion disks in broad-line emission, with particular emphasis on radio-loud quasars.  We applied a quantitative parametrization of the low-ionization broad emission line properties in the main sequence context. We stressed that broad emission lines show large red-ward asymmetry both in H$\beta$ and Mg II$\lambda$2800. An unbiased comparison matching black hole mass and Eddington ratio suggests that the most powerful RL quasars show the highest red-ward asymmetries in H$\beta$ in the general population of AGN. These shifts can be accounted for by gravitational and transverse redshift effects, especially for black hole masses larger than $  \approx 10^{8.7}$ M$_\odot$. The analysis of the extremely jetted quasar 3C 47 added another piece to the puzzle: not only are the low ionization profiles of 3C 47  well-described by a relativistic Keplerian accretion disk model, with line emission between $\approx$ 100 and $\approx$ 1,000  gravitational radii, but also the high-ionization line profiles can be understood as a combination of disk plus a failed wind contribution that is in turn hiding the disk emission. Constraints on radio properties and line profile variability suggest that 3C 47 might involve the presence of a second black hole with secondary-to-primary mass ratio $\lesssim 0.5$.  We conjecture that the  double peakers — type-1 AGN with Balmer line profiles consistent with accretion disk emission — might have their emission truncated by the sweeping effect of a second black hole. {  Our analysis of 3C 47 provides original evidence that the ubiquitous red asymmetry in Population B is consistent with gravitational and transverse redshift from the accretion disk.  In non-starving systems, the disk signal is plausibly masked  {by}  additional line emission, rendering the disk contribution harder to detect.}
\end{abstract}


\end{frontmatter}


\section{Introduction}
\label{intro}

Active Galactic Nuclei (AGN) are among the most luminous objects in the universe, powered by accretion onto supermassive black holes (SMBHs). In type-1 AGN, broad emission lines observed in the optical and ultraviolet spectra are thought to arise from gas in the broad-line region (BLR), located within $\sim 10^2 - 10^4$ gravitational radii from the central black hole \citep{mathewscapriotti85,osterbrockmathews86}. The BLR dynamics is still a debated issue. Some features are commonly attributed to the influence of a rotating, optically thick, geometrically thin accretion disk that contributes both to the ionizing continuum and, in some cases, to the broad-line profiles themselves. The exact role of accretion disks in producing these broad-line features remains unclear.

Observed broad-line profiles most often deviate significantly from the predictions of a simple rotating disk model \citep{sulentic89,sulenticetal90}. These deviations are pronounced for low-ionization lines such as \hb\ and \mgii, which exhibit diverse shapes, from symmetric Lorentzian to single-peaked, red-ward asymmetric, to multi-peaked profiles \citep[e.g., ][]{sulenticetal00a}. High-ionization lines, such as \civ, present additional complexities, with profiles influenced by outflows \citep[e.g., ][]{richardsetal06,sulenticetal07}. These features raise fundamental questions about the interplay between accretion disk physics, BLR dynamics, the role of relativistic effects, including gravitational redshift, transverse Doppler shifts, and radiative transfer phenomena. 

The main sequence of quasars \citep{borosongreen92,sulenticetal00a,sulenticetal00b,marzianietal01,shenho14,marzianietal18} provides a framework for organizing these diverse spectral properties, with powerful, relativistically jetted quasars (hereafter radio-loud, RL, for brevity) occupying distinct regions that may reflect differences in accretion mode, black hole spin, or jet coupling \citep{zamfiretal08}. RL  AGN offer a unique testbed for studying these processes due to their relativistic jets and often extreme physical conditions. Only systematic studies of broad-line properties within the main sequence context can effectively elucidate how these factors influence the emission-line profiles in both non-jetted and jetted systems. 

In the paper, we summarize and put in a broader context the recent results from an analysis of a large dataset of low-$z$ AGN \citep[][hereafter \citetalias{terefemengistueetal23}; see Section \ref{sample}]{terefemengistueetal23}, focusing on a major feature of the emission line profiles and on comparison between non-jetted and jetted sources.  The main results obtained by \citetalias{terefemengistueetal23} involved a quantitative analysis of the line profiles of the HI Balmer line \hb\ and of \mgii, and a discussion of the effect of powerful-relativistic jets on the AGN emission line properties (Section \ref{results}). We also summarize a study of the radio galaxy 3C 47 (\citealt{terefemengistueetal24}, hereafter \citetalias{terefemengistueetal24}), which exhibits Balmer and \mgii\ profiles consistent with emission originating from the atmosphere of an accretion disk. Finally, we explore the implications of this case for the broader AGN population (Section \ref{discussion}),   gaining new insight on the relation between the so-called “double peakers” (also known as ``disk emitters”) and the AGN belonging to Population B { in the quasar main sequence. }


\begin{figure}
  \centering
  \includegraphics[scale=0.5]{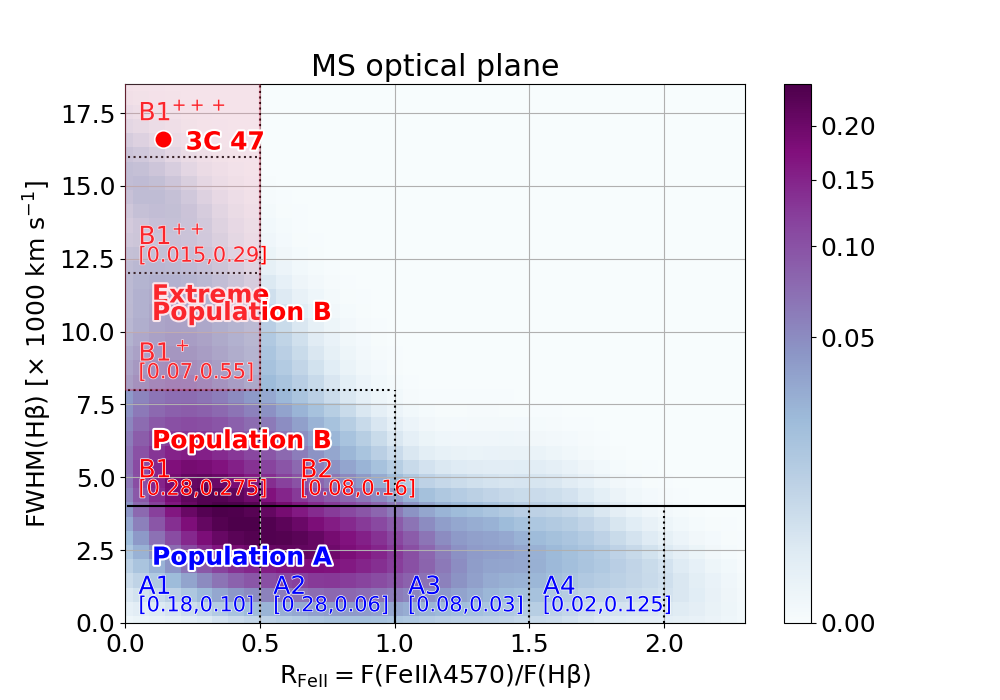}
  \caption{{  The quasar main sequence optical plane defined by the parameters \rfe\ and FWHM \hb, with a subdivision into Population A and B, extreme Population B, and spectral types. The numbers below each spectral type identification are the prevalence of AGN {in each spectral type,} followed by the prevalence of radio-intermediate and RL AGN in each spectral bin. The shading is proportional to the surface density of AGN in the sample of \citet{zamfiretal10}, and has been expanded at low density values to better represent of sparsely occupied spectral bins. The location of 3C 47 in the main sequence  (discussed in Section \ref{3c47}) is at an extreme of extreme Population B (the area shaded rose encompassing the B1${^{++}}$ and B1$^{+++}$\ spectral types).  }}
  \label{fig:msprof}
\end{figure}

\section{Samples and methodology}
\label{sample}
 \subsection{Samples} 
 
\citetalias{terefemengistueetal23} joined the sample of \citet{marzianietal03a} and the SDSS-based sample of \citet{zamfiretal10} and obtained a sample of 640 low redshift AGN in the range of redshift $0 \lesssim z \lesssim 1$ and bolometric luminosity $10^{43} - 10^{47} $ \ergs. Optical and near-UV spectra of 11 powerful jetted quasars in the redshift range $0.35 < z < 1$, observed with the Cassegrain TWIN spectrograph of the 3.5 m telescope at the Calar Alto Observatory (CAHA) in Spain, were added to this sample,  {as described in}  \citetalias{terefemengistueetal23}. Their radio-to-optical flux density ratio  {of}  $> 10^3$ is extreme.   The  joint sample encompasses 651 type 1 AGN with spectral resolution $\lambda/\delta\lambda \sim 1,000 - 2,000$, {and} continuum S/N $\gtrsim 20$  \citepalias[see][]{terefemengistueetal23}. Resolution and S/N are sufficient to perform a careful deblending between components of the same line, or by other superimposed emission features (e.g.,  \feii\ contaminating \hb).  We distinguish between radio-quiet (RQ), radio-intermediate, and RL sources on the basis of a slight modification of a criterion originally defined by \citet{kellermannetal89},  using the radio-to-optical flux ratio of  $R_{\mathrm{K}}= f_\nu$(1.5GHz)$/g < 10$, $10 \le R_{\mathrm{K}} < 63$, and $R_{\mathrm{K}}\ge 63$\, respectively. These criteria are intended to serve as a sufficient ({  albeit not necessary}) condition for the selection of powerful, relativistically jetted sources classified as RL, while excluding  radio-intermediate sources which  are  more heterogeneous \citep{gancietal19}.  

 {  Figure \ref{fig:msprof} shows  the occupation}  of the optical plane of the quasar main sequence  \rfe\ using the sample of \cite{zamfiretal08}. {The prevalence of AGN in each spectral bin is reported below the spectral type identification, followed by the prevalence of AGN with $\log R_\mathrm{K} \ge 10$.} {  The values reported in Fig.  \ref{fig:msprof} indicate} that the prevalence of jetted AGN is higher in Population B, {  by a factor $\approx 3 - 5$, with a surge in bin B1$^{+}$, which however involves  only $\approx $ 7\% of all AGN \citep{gancietal19}. Different optically selected samples (e.g., { \citealt{marzianietal13a,marzianietal13})} may yield a slightly different prevalence. Nonetheless, the same trend is observed: a higher incidence of powerful radio sources in Population B, along with an increasing RL fraction with Balmer line full width at half maximum (FWHM) \citep[e.g.,][]{chakraboryetal22}.} Even if these numbers are affected by a Malmquist bias due to an (admittedly ragged) flux limit, they reflect the rarity of jetted sources toward the opposite end of the sequence. RL Narrow Line Seyfert 1s \citep[NLSy1s,][]{komossaetal06,foschinietal15} are a rather rare phenomenon   \citep{komossaetal06,paliyaetal24}, and  no source meets the criterion $\log R_\mathrm{K} \ge 1.8$\ out of the 91 NLSy1s of the joint sample.  
 
\subsection{Line profile analysis}

The \hb\ profiles of composite \hb\ spectra of Population  B show a prominent red-ward asymmetry ({  left panel} of Fig. \ref{fig:histc14}) that is not observed in Population A composites,  \citep[{  see for example, }][] {sulenticetal02,marzianietal10,negreteetal18}. The \citetalias{terefemengistueetal23} analysis includes the standard spectrophotometric measurements (e.g., FWHM, continuum and line fluxes)   and derived quantities (luminosity, black hole mass \mbh, Eddington ratio \lledd). In addition, centroids at different fractional intensity (Fig. \ref{fig:histc14}) provide a quantitative characterization of the line profiles. The centroids are defined with respect to the quasar rest frame, in practice provided by the narrow component of \hb. The centroid at one-quarter intensity $c(\frac{1}{4})$ is used as a measure of the line asymmetry {  toward the line base}. 

\subsection{Contextualization along the main sequence of quasars}

The quasar main sequence categorizes type-1 AGN based on the correlation between the strength of \feii$\lambda$\ 4570 blend and the FWHM of H$\beta$.  It highlights the presence of two main populations: Population A and Population B, separated at FWHM $\approx$ 4000 \kms, in moderate luminosity samples ($\log L \sim 46$ \ergs).  Population A AGN  have  a narrower \hb\ line  (FWHM $<$ 4000 \kms),  a range of \feii\ emission  (from weak to high \( R_\mathrm{FeII} \)), and often lower \civ\ and \oiii\  equivalent width \citep{borosongreen92,sulenticetal02,dultzin-hacyanetal97}. Population A quasars are typically associated with higher Eddington ratios    \lledd $ \approx 0.2 - 1.0$, compared to the lower values of $L/L_{\mathrm{Edd}} \approx 0.01 - 0.2$ in Population B \citep{marzianietal01,marzianietal03b,pandaetal19}. They are overwhelmingly RQ.  Population B quasars have  broad \hb\ lines  (FWHM $>$ 4000 \kms),  weak \feii\ emission  (low \( R_\mathrm{FeII} \)), and stronger \oiii\  emission \citep{borosongreen92,sulenticetal02,marzianietal03a}. Population B quasars are associated with lower accretion rates, and higher \mbh, typically ranging from $10^{8.5}$ to $10^{9.5} M_{\odot}$ \citep[][\citetalias{terefemengistueetal23}]{marzianietal03b,marziani23}. Even if there is no obvious boundary between Population A and B in the data point distribution, at FWHM $\approx $ 4000 \kms\ we observe a change in multi-frequency properties {  between Population A and B}   (for an exhaustive summary, see  \citealt{sulenticetal11}, \citealt{fraix-burnetetal17} and \citealt{marzianietal18}). 


\section{Results}
\label{results}

\begin{figure}
  \centering
  \includegraphics[scale=0.55]{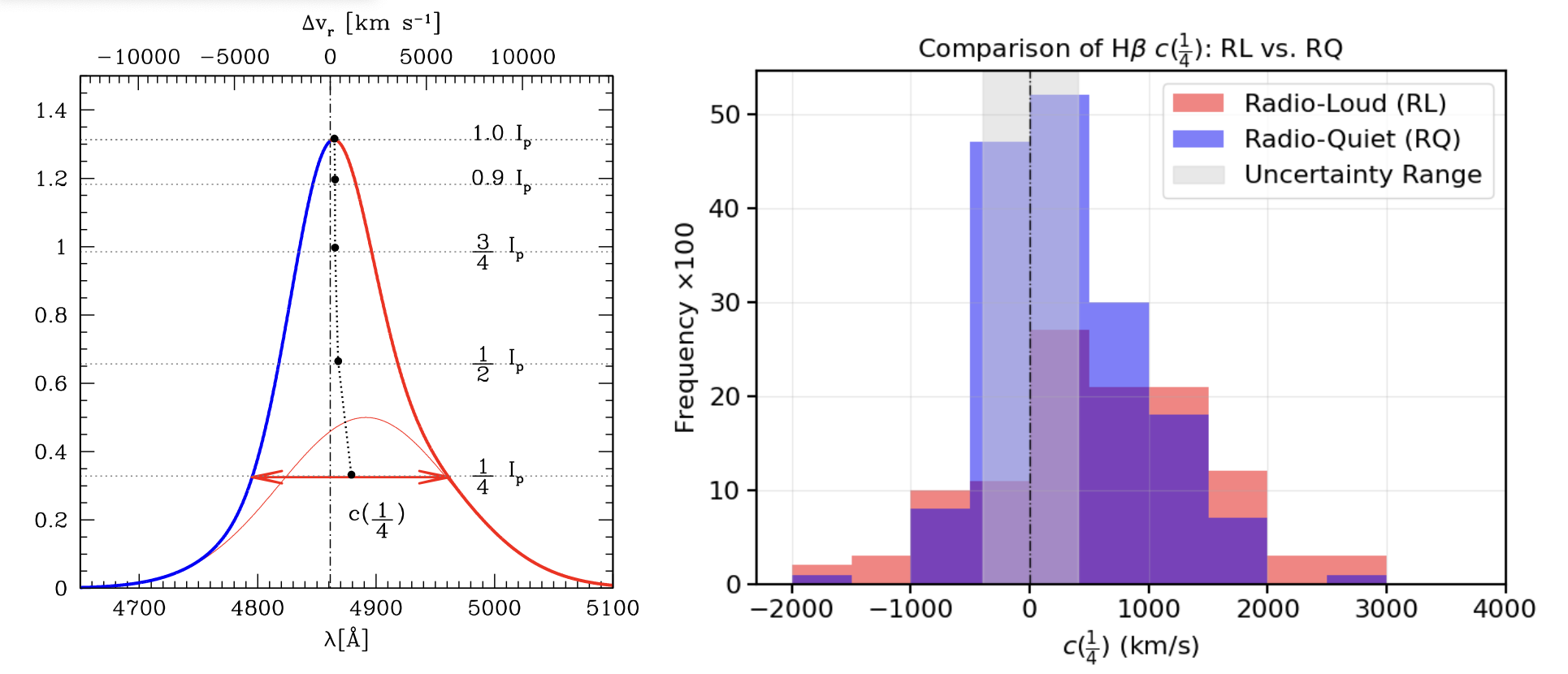}
  \caption{{  Left panel: sketch of a typical Balmer line profile (\hb) of a Population B AGN 
  with the centroids at different fractional intensity marked, with an emphasis on the $c(\frac{1}{4})$\ analyzed in this work. The thin red line traces the putative ``very broad component" that might ultimately be {  traced back} to accretion disk emission, and is represented by a red-shifted Gaussian of FWHM $\sim 10,000$ \kms.   Right:}  Distribution of H$\beta$ centroid $c(\frac{1}{4})$ for Population B RQ (blue) and RL (red), {  for the sample of \citet{zamfiretal10}}. The shaded area is the uncertainty range at a $\pm 2 \sigma$\ confidence level.}
  \label{fig:histc14}
\end{figure}

\subsection{Population A  and Population B}	

The analysis of the joint sample confirms several of the trends and systematic differences between Population A and B that were already derived in the earlier works (\citealt{marzianietal03a,zamfiretal10}; see \citetalias{terefemengistueetal23} for a full report). We just point out that line asymmetries in \hb\ are systematically different between the two Populations, with a much larger prevalence of red asymmetries (positive $c(\frac{1}{4})$ values) among Population B sources, and that $\approx 80$\%\  of the RL subsample (157 sources) belongs to Population B ({see Fig. \ref{fig:histc14})}.  

\subsection{Population B jetted and non-jetted AGN}

However, a proper comparison should be carried out, restricting the samples to Population B, to which 50\% \ of the RQ sources of the joint sample also belong. The differences are found to be more elusive when the comparison is carried out by restricting the RQ population to the region of the quasar main sequence occupied by RL sources (i.e., Population B, consistent domains in black hole mass and Eddington ratio): the RL and RQ quasars spectral properties become more similar, also in terms of the strong red-ward asymmetries in H$\beta$\  emission lines. Fig. \ref{fig:histc14}  (right panel)  shows the distribution of the \hb\ $c(\frac{1}{4})$ for Population B and RQ and RL sources: both RQ and RL are skewed toward   the red.  A bootstrap analysis, matching distribution of \lledd\ and \mbh, confirms that there is no significant difference between RQ and RL \citepalias{terefemengistueetal23}. Only the most powerful RL quasars display systematically stronger red-ward asymmetries in the H$\beta$\  emission lines with respect to the matching RQ sample.  

\begin{figure}
  \centering
  \includegraphics[scale=0.41]{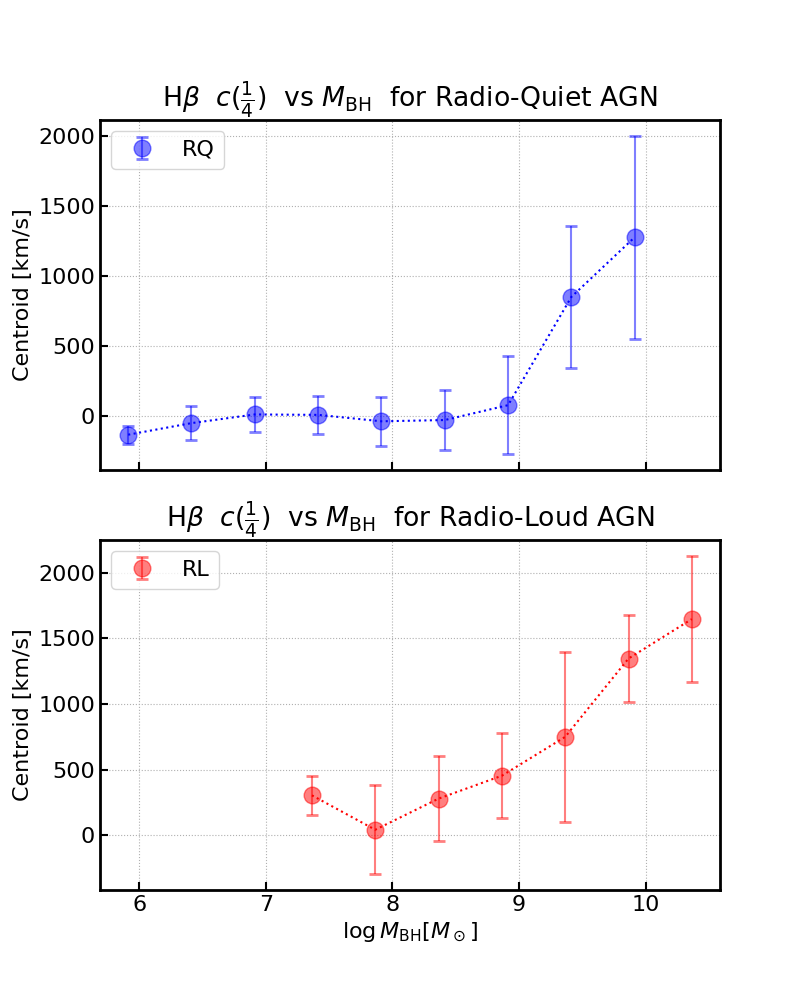}
  \caption{Behavior of centroid at $\frac{1}{4}$\  peak line intensity of \hb\ as a function of black hole mass, for RQ (top) and RL (bottom) AGN of the \citetalias{terefemengistueetal23} sample. Data points are median centroid values in ranges of 0.5 dex; error bars are sample semi-interquartile ranges.   
  \label{fig:cmass}}
\end{figure}

Fig. \ref{fig:cmass} illustrates the behavior of the  \hb\ centroid at $\frac{1}{4}$\ as a function of black hole mass. Data points are medians in \mbh\ intervals of 0.5 dex, and error bars are the corresponding semi-interquartile ranges. The $c(\frac{1}{4})$\ parameter in Fig. \ref{fig:cmass} is a  measurement of the extension of the red wing.  The profiles are symmetric for low \mbh, where the Population A is dominating the medians. Low mass RL are absent from the sample. In the optically selected samples under consideration, there is a threshold for radio-loudness, as RL sources with \mbh \ $\lesssim 10^{{7.5}}$\ M$_{\odot}$\ are absent (Fig. \ref{fig:cmass}).

For both RQ and RL we see a consistent increase in $c(\frac{1}{4})$\ above \mbh $ \sim 10^{8.5}$ M$_\odot$ (Fig. \ref{fig:cmass}).  In summary, RL quasars may exhibit some extreme values in terms of red-ward asymmetry and mass, but the distribution of most of them overlaps with the one of the RQ. This points toward {  no major}  differences in the history of black hole growth between the two populations, rather than toward an effect of the relativistic jet: the phylogeny of the central black hole sets the ontogeny of the quasar populations \citep[][\citetalias{terefemengistueetal23}]{fraix-burnetetal17}.

\subsection{Origin of the red-ward asymmetry}

The origin of the red-ward asymmetry has been debated since the early 1970s. The increasing redshift toward the line base has been explained as due to gravitational redshift \citep{netzer77,corbin95,gavrilovicetal07,marzianietal09,bonetal15,marziani23}. The appeal of gravitational and transverse redshift resides in the fact that the red-ward asymmetry is usually a very stable feature, and widespread involving Population B sources that are about $50$ \%\ of optically selected samples. Alternative scenarios could involve infall and obscuration of the receding part of the flow, possibly coexisting with virial motion \citep{gaskell10a,baraietal12,wangetal13,lietal22}, or anisotropy in the line emission \citep{ferlandetal09}. Other explanations invoked off-axis emission that violates the assumption of a perfect axial symmetry of the BLR \citep{gaskell10b,storchi-bergmannetal93,storchi-bergmannetal03}, gravitational recoil of a black hole following coalescence \citep{favataetal04,merrittmilosavljevic05}, or a binary BLR, with each BLR bound to an individual black hole of the pair \citep{gaskell84}. In general, these models assume configurations that are time-evolving, and eventual monitoring did not provide supporting evidence.   

  \begin{figure}
  \centering
  \includegraphics[scale=0.25]{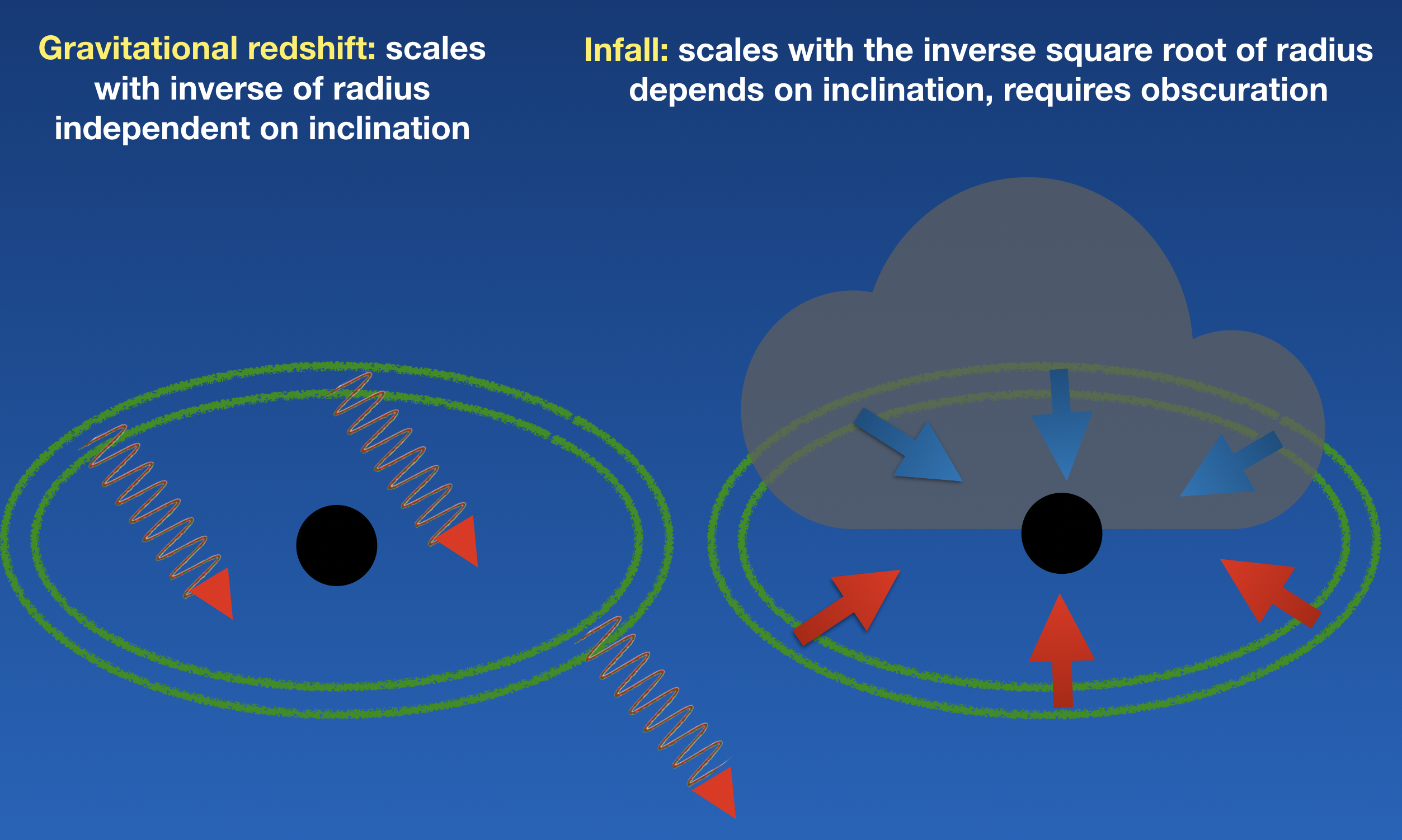}
  \caption{Sketch illustrating the differences expected for gravitational redshift   (left) and redshift due to gas with an infalling component of its velocity where the approaching side of  the inflow is hidden/obscured (right). The photon gravitational redshift from an annulus at a given distance 
 from the central black hole is independent of the velocity fields. In contrast, in the case of infall, the blue and red colors of the arrows represent the Doppler shifts (blue-shift and red-shift) caused by the radial component of the gas velocity field. The gray cloud indicates that the approaching side of the gas yielding a blueshift must be obscured to produce a net redshift in the emission line profile. }
  \label{fig:gravzinf}
\end{figure}

Fig. \ref{fig:gravzinf} sketches an idealized representation of the effect of gravitational redshift and infall. An annulus at a given distance from the central black holes emits photons that are shifted because of the black hole gravitational field; a transverse redshift component is due to high velocity of the line emitting gas, yielding a total shift to the red $z \sim  \frac{3}{2} GM_\mathrm{BH} /c^{2}r = \frac{3}{2} r/r_{\mathrm{g}}$.  The case of infall requires some more ad-hoc assumptions, such as that the approaching side of the inflow is obscured, as schematically shown in Fig. \ref{fig:gravzinf}. Inflow of gas close to the equatorial plane is opposed by radiation forces; it is reasonable that the infall velocity will be a fraction of the free fall velocity, $v_\mathrm{inf} \lesssim  \sqrt{GM_\mathrm{BH} /r}$. The two cases predict different trends with the black hole mass: $\propto M_\mathrm{BH}$\ for gravitational redshift, and $\propto  \sqrt{M_{\mathrm{BH}}}$.  If we include the dependence on the radius (roughly proportional to $r^{\frac{1}{2}}$, we expect a slope between $c(\frac{1}{4})$\ and \mbh\  $\approx 0.5$ and  $\approx 0.25$ for gravitational redshift and infall, respectively.  The $c(\frac{1}{4})$\ is indeed significanty  correlated with \mbh, and the slope is $\approx 0.5$, consistent with the gravitational redshift hypothesis  if \mbh $\gtrsim 5 \cdot 10^{8}  M_{\odot}$ \citep{marziani23}. The effect is  visible in Fig. \ref{fig:cmass}, where the median values of  $c(\frac{1}{4})$\ increase with a slope $\approx 0.5$\ as a function of mass, for both RQ and RL sources.  The red-ward asymmetry might be  ascribed to gravitational redshift, at least in systems with very large black hole mass.  

\subsection{Double peakers: the bare accretion disk?}

Double-peaked emission lines are a remarkable feature observed in the spectra of certain AGN. These features, particularly prominent in low-ionization lines such as \hb\  and \mgii, are rare \citep{stratevaetal03,fuetal23} and are widely interpreted as signatures of gas rotating in a Keplerian accretion disk surrounding a supermassive black hole \citep{chenhalpern89,eracleoushalpern03}: a structure above the inner disk  illuminates the outer disk and drives the
line emission \citep{chenetal89,dumontetal90,riccisteiner19}. 

The distinctive double-peaked profiles arise from the relativistic Doppler effect, with the red and blue peaks corresponding to emission from opposite sides of the disk. Relativistic effects, while minor, are not negligible: the blue peak appears stronger than the red due to Doppler boosting of radiation from the approaching gas. Meanwhile, the red wing is shifted further red-ward, reflecting the combined influence of the virial velocity field and the increasing gravitational and transverse redshift as the gas approaches closer to the black hole. 

Double peakers are located in correspondence of spectral types B1$^{++}$ or B1$^{+++}$ along the main sequence, which encompass the broadest emission lines but involve  only a few percents of the SDSS AGN \citep{zamfiretal10,shenetal11}.   Double-peaked emission lines are more commonly observed in RL AGN, though they are not exclusive to this class \citep{stratevaetal03}. The high prevalence among RL does not suggest a fundamental connection between the radio jets and the structure or dynamics of the accretion disk, but rather, a pattern in the growth of the supermassive black hole similar to the one of RQ AGN with consistent mass and Eddington ratio \citep{marzianietal22a}.  Intriguingly, the red wing of double peakers behaves as expected if gravitational redshift is affecting the line emission — a circumstance that hints at the possibility that an accretion disk contribution might be common in Population B AGN \citep{popovicetal04,bonetal09a,storchi-bergmannetal17}.  It remains to be seen why some — definitely not all — sources in the extreme Population B show the double peaked structure in their low-ionization lines. 

\subsection{3C 47: a case study and its implications}
\label{3c47}

3C 47, a RL quasar with an optical redshift of $z \approx$ 0.4248, serves as an exemplary case for studying accretion disk and BLR dynamics. It exhibits extreme properties such as high black hole mass and extreme radio loudness ($\log R_\mathrm{K} \approx 4$). The \mbh\ and \lledd, equal to $(7 \pm 1) \times 10^{9}$ M$_\odot$\ and 0.04$\pm0.01$  are consistent with the B1$^{{+++}}$\ spectral type (borderline B1$^{{++}}$) \ classification inferred from the line width (FWHM \hb\ $\approx 16,600 $ \kms; Fig. \ref{fig:msprof}). 

 A relativistic Keplerian accretion disk model computed by \citetalias{terefemengistueetal24} successfully explains the observed low-ionization line double-peaked profiles (H$\beta$, and H$\alpha$), near-UV (\mgii). Bayesian inference with uniform and Gaussian priors has provided tight constraints on the accretion disk parameters for 3C 47. A modified Metropolis-Hastings MCMC  fit using relativistic disk models  following \citet{chenhalpern89} yielded the values of parameters such as inner radius ($r_\mathrm{in} \sim$ 100 $r_{\mathrm{g}}$), outer radius ($r_\mathrm{out} \sim$ 1,000 $r_{\mathrm{g}}$), and inclination angle  ($\theta \approx 30$) which are fairly typical for double peakers \citep{stratevaetal03}.  
 
An alternative fit, employing two Gaussian component symmetrically displaced in radial velocity with respect to the rest frame gave a similar $\chi^{2}$. The underlying physical scenario would be a binary BLR, with two BLR each associated with a black hole rotating around the common center of mass of the binary \citep{gaskell84,halpernfilippenko88,eracleousetal12,popovic12,doraziocharisi23,komossagrupe24}. The recent discovery of a gravitational wave background by pulsar-timing arrays points to a significant population of massive black hole binaries \citep[e.g.,][and references therein]{chenetal24a,verbiestetal24}. The presence of two peaks would require the absence of circum-binary emission \citep{popovicetal21}, which is likely in a starving system like the one of 3C 47 \citep{fabiancanizares88,ho08}. The binary BLR models imply that the peak of the two components should displace on the timescale of the binary black hole orbital period. The actual displacement of the peaks (or at least an upper limit to it) gives in turn a strong constraint on the sum of the binary black hole mass \citep{eracleousetal97}:  a very long period implied by a small or uncertain radial velocity displacement may imply in turn  unrealistically  large \mbh. This seems to be the case for most, if not all, double peakers \citep{halperneracleous00,breidingetal21,doanetal20,runnoeetal15,runnoeetal17}. In the case of 3C 47, coarse upper limits to the \hb\ and \mgii\ peak displacement already imply a black hole mass that is several times larger than $10^{10}$ M $_{\odot}$, a condition that is rare \citep{sulenticetal04,marzianisulentic12,trakhtenbrotnetzer12}, even if not theoretically impossible \citep{natarajantreister09,king16}.  

At the same time, not all double-peaked profiles are immediately consistent with the disk profile with mild  {relativistic} effects. Different relative intensities of the peaks might require a configuration that is not axisymmetric, or affected by additional components such as the wind from the accretion disk itself \citep{eracleousetal95,flohicetal12}. While the 2012 observations of 3C 47 showed \hb, H$\alpha$, and \mgii\ profiles are in very good agreement with the relativistic accretion disk model, spectra obtained at earlier epochs are too sparse to reach any firm conclusion apart from disfavoring the binary BLR explanation.  

\begin{figure}
  \centering
 \includegraphics[scale=0.41]{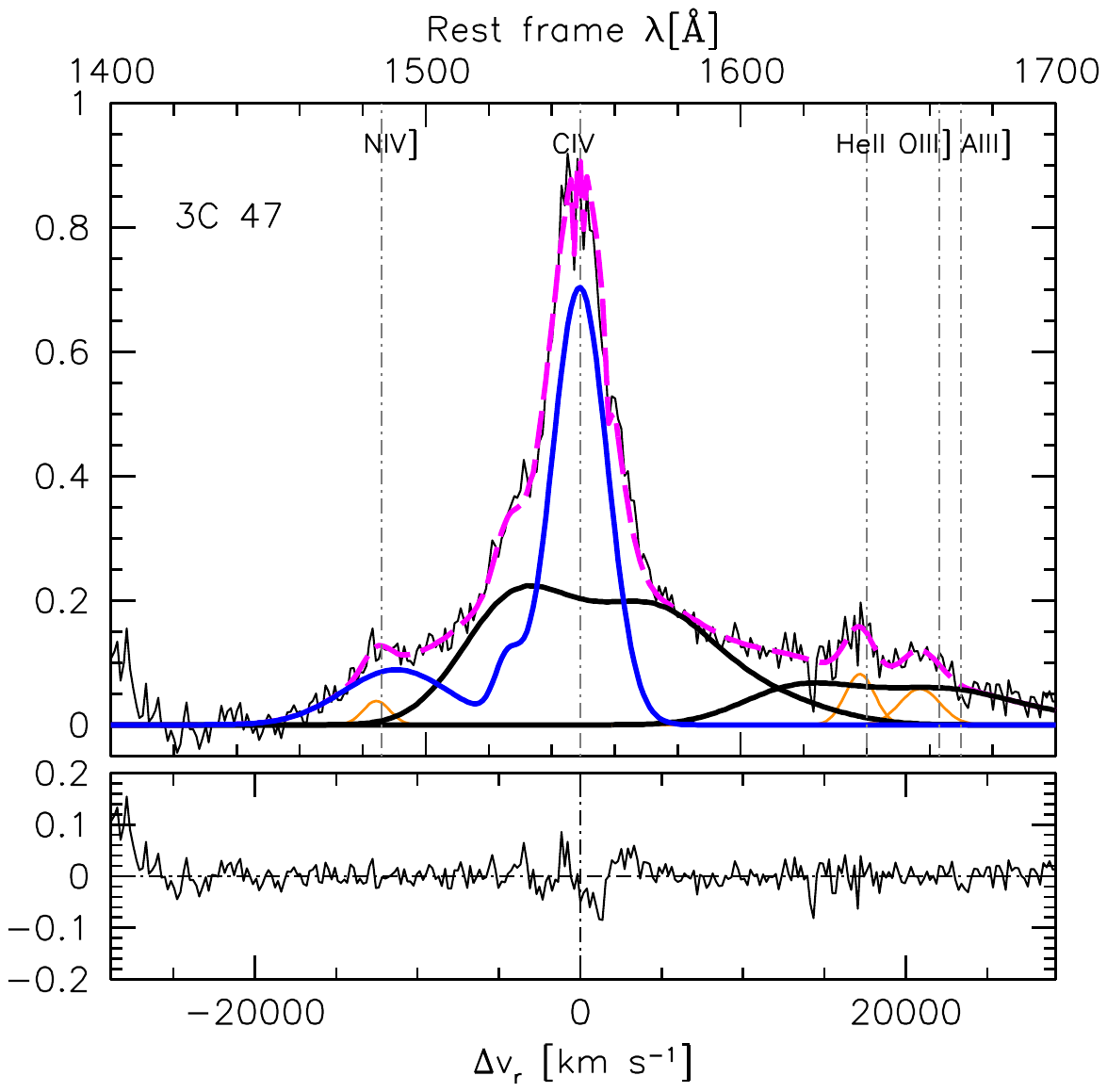}
 \vspace{-2cm}
  \caption{Interpretation of the \civ\ line blend. The thick black line is the accretion disk profile as derived from the \hb\ fit, scaled to the radial velocities of \civ\ and \heiiuv. The blue thick line represents the failed wind emission, in the form of a strong, symmetric line component plus a faint wing whose shape is highly uncertain. Orange line represent semi-broad components of \heiiuv, {\sc O  iii}]$\lambda 1663$,  {\sc Niv}]$\lambda 1486$, {\sc Alii}]$\lambda 1670$. Ordinate is specific flux in arbitrary units. The bump on the blue side of \civ\ at $\approx -12,000$ \kms\ is likely associated with the outflow, as the intensity of {\sc Niv]}$\lambda 1486$ is expected to be just a few hundreds the one of \civ\ from photoionization simulations. The narrow component of the \civ\ profile is suppressed by narrow absorption lines clustered close to the rest-frame of the quasar. Adapted from { {Fig. 8 in}} \citetalias{terefemengistueetal24}.}
  \label{fig:chrev}
\end{figure}

\begin{figure}
  \centering
 \includegraphics[scale=0.41]{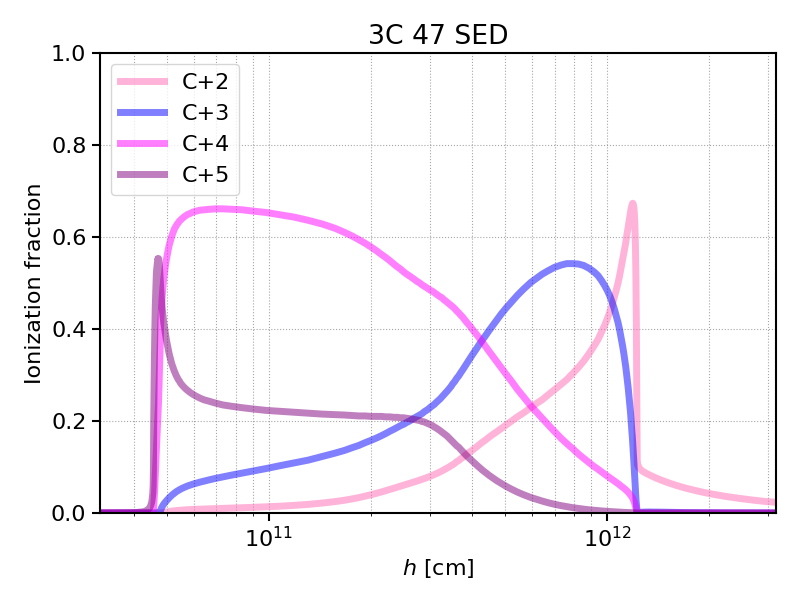}
 \includegraphics[scale=0.41]{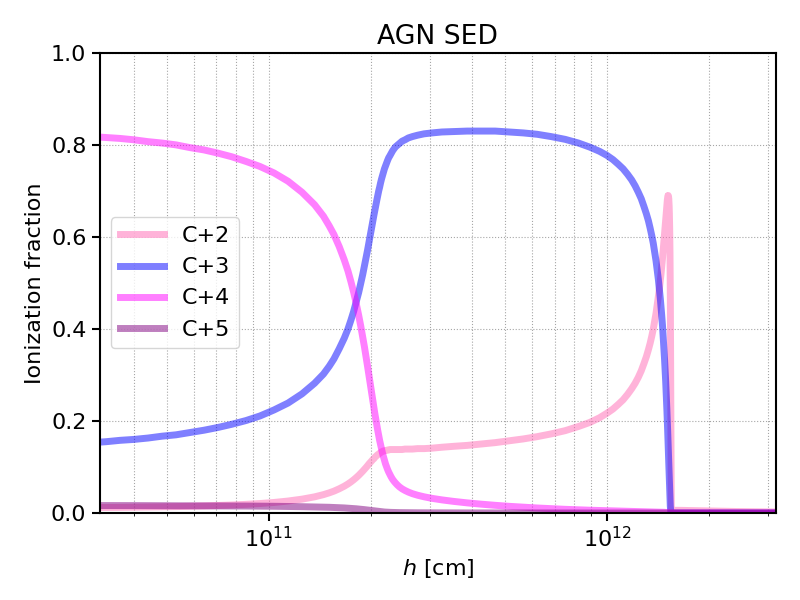}
  \caption{Ionization structure as a function of the depth within a slab of emitting gas, for photoionization by the 3C 47 SED (left) and a standard AGN SED \citep{mathewsferland87}. Only the most populated ionic species of carbon are shown.  {  Adapted from Fig. 10 in \citetalias{terefemengistueetal24}. For more information see the text and \citetalias{terefemengistueetal24}.  } }
  \label{fig:ion}
\end{figure}

\subsubsection{Accretion Disk and Outflow Models}
 
The analysis of 3C 47 extended to lines over a broad range of ionic species suggests a complex interplay between virial motion and radiatively driven outflows. The profiles of the prototypical high- and intermediate-ionization UV lines (\civ, \ciii)  can also be modelled by an accretion disk contribution, but require a fairly symmetric additional components superimposed to the accretion disk profile (see Fig. \ref{fig:chrev}, and Figs. 8 and 9 of  \citetalias{terefemengistueetal24}). The appearance of the \civ\ line is single-peaked, with a FWHM $\sim 4000$\kms\ much lower than that   of \hb\ and \mgii. We note in passing that this situation is in stark contrast with respect to {\textcolor{red}{the}} source at the opposite end of the main sequence, where we encounter that the \civ\ line is broadened by blueshifted component due to a powerful outflow, which makes \hb\ significantly narrower than \civ\ \citep{sulenticetal00a,leighlymoore04,sulenticetal07,coatmanetal16}. 

Returning to Population B, AGN in this category -- particularly RL Population B sources -- exhibit a spectral energy distribution (SED) with significant emission in the X-ray domain and a less prominent big blue bump compared to Population A sources  \citep{laoretal97,marzianietal21a,pandamarziani23}. This SED is poised to create a condition of over-ionization, if the line-emitting gas is fully exposed to it. Fig. \ref{fig:ion} shows the ionization structure within a gas slab with a column density $N_{\mathrm{H}} = 10^{23}$ cm$^{-2}$, hydrogen density $10^{10}$ cm$^{-3}$, and distance from the continuum sources $\log r = 18$\ [cm] corresponding to $\approx $1,000 $r_{\mathrm{g}}$, where the illuminated surface is at a depth $h = 0$ (left side {  of the left panel}), for the continuum of 3C 47 (left panel), and for a typical AGN continuum normalized to the same optical luminosity (right panel). The Figures show the main ionic species of carbon: in the case of the 3C 47 SED, ionic stages higher than C$^{+3}$ dominate the emission close to the slab face exposed to the continuum, and the ionic stage C$^{+3}$\ which permits the emission of \civ\ reaches only $\approx$0.55 of all carbon atoms. In contrast, a standard AGN SED \citep{mathewsferland87} produces a higher fraction of C$^{+3}$\ over a larger extent in the cloud. As a consequence of  over-ionization, the force multiplier is  lower for 3C 47, and a wind is accelerated less efficiently \citep[][c.f. \citealt{czernyetal17,naddafetal21a}]{murraychiang98,progaetal00,proga07}.   The result is a symmetric, mostly virialized \civ\ component which is interpreted as due to a “failed wind,” between the outer radius of the accretion disk and the innermost part of the narrow line region.   

Another piece of the puzzle is added by the ``disappearance'' of the narrow component of \civ\ \citep{willsetal93,sulenticmarziani99}: the \civ\ profile apparently is topped by a narrower core, but with FWHM 
 $\sim 1,000 - 2,000$ \kms\ and so much broader than the narrow component of \hb, typically from a few to several hundred \kms\ in Population B \citep{sulenticetal00a}.
The continuity with the \civ\ NC is due to fact that the \civ\ emissivity, in   dust-free conditions, is heavily weighted in favor of the higher density and ionization parameter innermost region, while the total, spatially unresolved emission of \hb\ and \oiii, with the same distribution of emitting gas, is due mostly to the outer regions \citep{sulenticmarziani99}.  This creates a characteristic discontinuity in the profile of \hb\ between the narrow- and broad component, which is absent in \civ.  In 3C 47, \civ\ NC is highly uncertain because of the contamination by narrow absorption lines clustered around its rest frame. 

\begin{figure}
  \centering
  \includegraphics[scale=0.41]{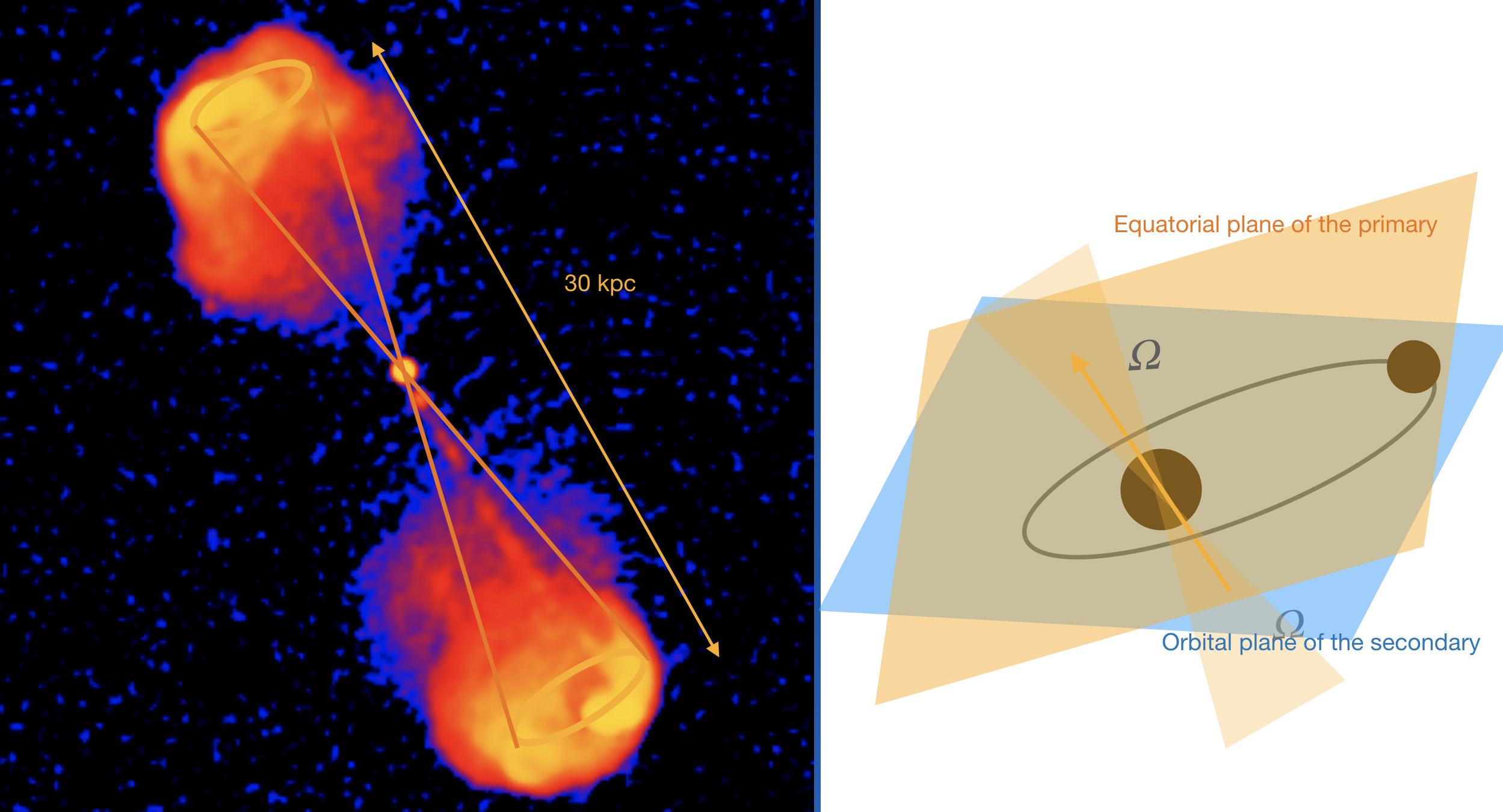}
  \caption{Left: VLA map of 3C 47 obtained on 07/12/92 at 18cm and retrieved from the NASA/IPAC Extragalactic Database (NED). A possible pattern of precession for the radio jet is overlaid on the map. Right: a sketch depicting the presence of a secondary black hole whose orbit is inclined with respect to the equatorial plane of the primary. }

  \label{fig:jetprec}
\end{figure}

Is this the end of the binary black hole story? No. The evidence of a second black hole comes from the radio morphology of 3C 47, shown in Fig. \ref{fig:jetprec} (1), a jet pointing toward the edge of southern lobe, where (2) a ring-like feature  is well visible   \citep{ferninietal91,leahy96}.  (3)  A third evidence is provided by the S-symmetry of the hot spots with respect to the jet axis.  The morphology suggests a precession of the jet axis describing a cone of full aperture around $ 25 $ degrees (Fig. \ref{fig:jetprec}){ : the} most straightforward interpretation is that a precessional motion is induced by a second black hole, with expected period $P_\mathrm{p} \sim 124 (1 + q)^{2}/[q(3q + 4)] r^\frac{5}{2}_\mathrm{pc} M_{9}^{-\frac{3}{2}}$ Myr \citep{krauseetal19}. Considering that the \mbh\ is $\approx 7 \cdot 10^{9}$ M$_{\odot}$, and that a second black hole orbit cannot reach a periastron less than 1,000 $r_{\mathrm{g}} \approx 0.34$ pc, the geodetic precession period can always exceed the  dynamical timescale of the radio source  ($\tau_\mathrm{d} \approx 6 \times 10^{7}$ yr) if $r \approx 10,000  r_{\mathrm{g}}$ i.e., the second black hole is located toward the inner edge of the torus (which may not be present in the starving conditions of 3C 47). A precession timescale significantly longer than the dynamical timescale of the radio source would permit to maintain an ordered appearance of the morphology, and this condition is met, for example,  if $q \sim 0.1$. This $q$ value would yield a precession period  {  $\sim 4 \times 10^{8}$ yr.} A higher $q$\ would be possible if  $r$\ is large enough to obtain  $P_\mathrm{p} \gg \tau_d$.  

\section{Discussion}
\label{discussion}

The \civ\ and the low-ionization lines profiles of 3C 47 provided a sort of  rosetta stone in the interpretation of the relation between wind and disk, at least in the case for very massive black hole radiating at very low Eddington ratio, close to the limit of radiative efficiency \citep{giustiniproga19}. All FRII, type-1 AGN like 3C 47 of high power are all high-excitation radio galaxies (HEG) in terms of their narrow line emission \citep{buttiglioneetal10,buttiglioneetal11}, supporting their interpretation as due to supermassive black hole accreting in the radiatively-efficient domain. 

The significance of 3C 47 in this context lies in providing a direct clue for interpreting the line profiles: a clear and easily identifiable contribution from the accretion disk in the \civ\ profile shown in Fig. 8  of \citetalias[][]{terefemengistueetal24} (c.f. \citealt{halpernetal96}). The low-ionization, virialized part of the BLR, given the typical metallicity of AGN in Population B (solar or slightly subsolar, but at any rate $\gtrsim 0.1 $ Z$_{\odot}$, \citealt{marzianietal23,marzianietal24,florisetal24}), is expected to show significant \civ\ emission, with ratios \civ/\hb\, $\gtrsim 1 $. The failed wind component, with prominent \civ, and faint \hb, indicates \civ/\hb\, $\gg  1$, as the emitting gas is probably exposed to the full AGN continuum, coming from the thermal continuum of the accretion flow and from the synchrotron-self Compton X-ray component \citep{ginzburgsyrovatskii69,konigl81,maraschietal92}.   As the source becomes less ``starving,'' more gas over a broader range of distance is illuminated by a stronger thermal continuum, giving the \hb\ emission line profile a single peaked appearance while leaving the failed wind configuration basically unaltered, unless the accretion rate increases to \lledd $\gtrsim 0.1 - 0.2$, and a source enters in the Population A domain. A quantitative comparison between \civ\ and \hb\ line profiles shows that powerful RL sources have rather similar profiles \citep{marzianietal96,corbinboroson96}, with the red asymmetry visible also in \civ\ and \heiiuv, decomposed into a broad and a very broad component, as for \hb\ (Fig. \ref{fig:histc14}), but with the addition of a faint blue-shifted component associated with a minor outflow component \citep{fineetal10,marzianietal10}. At least in some objects of Population B, the very broad component in the mock \hb\ profile of   Fig. \ref{fig:histc14} could be ascribed to an accretion disk contribution \citep{popovicetal04,bonetal09a,storchi-bergmannetal17}. The similarity of  the \civ\ and \hb\  full  profiles is accounted for because both lines are mostly the sum of two virialized components. 

\subsection{Starvation and/or truncation}

The supermassive black hole of 3C 47 is accreting at very low rates, as is the prototypical ``double peaker'' Arp 102B. Starvation appears as one of the necessary conditions for the existence of double speakers, which also need to be observed at a rather large angle between the line of sight and the disk axis.   Limits on periodicity of Balmer line variations and radio morphology hinting at precession induced by a second black hole in the case of 3C 47 suggest that the minority class of double peakers might be due a disk truncation by a second black hole orbiting the massive primary.  It is interesting to note that a large fraction of FRII galaxies show morphological evidence in the radio band that suggest jet precession \citep{krauseetal19}. FRII sources are over-represented in the spectral type B1$^{++}$, accounting for a large fraction ($\lesssim 30 $\%) of all sources in this spectral type \citep{gancietal19}. The prototypical double peaker Arp 102B shows an excess of broad emission in correspondence of the rest frame, over the profile of the accretion disk model. If emission is predominantly virialized, this may imply a discontinuity in the line width and hence a gap in the range of disk radii where line emission is produced, as expected in the case of a second black hole \citep[e.g.,][]{liushapiro10}. Alternatively, analytical models and numerical simulations of gas flows around orbiting binaries suggest that a centrifugal barrier ought to 
inhibit mass transfer onto the binary \citep{tiedeetal22}. In this case, however, two mini disks are expected to form around each black hole, and the variation of the profiles should still be associated with orbital motion. The whole double-peaked profile of the accretion disk of the more massive black hole should show {\textcolor{red}{a}} periodic shift proportional to the mass ratio $q$. 

\section{Conclusion}

The analysis presented in this paper has been focused on two lines of investigation, following and expanding on three recent works \citep[][\citetalias{terefemengistueetal23,terefemengistueetal24}]{marziani23}. The first one was centered on the broad emission low-ionization line profiles (\hb\ and \mgii) of RQ and RL quasars within the main sequence framework, quantifying their red-ward asymmetries and their dependence on black hole mass. The second one involved the analysis of the jetted quasar 3C 47, whose emission-line properties provide a prototypical example of accretion disk-dominated low-ionization line profiles. The {  low-and high ionization lines}  of 3C 47 offer insights into how relativistic accretion disk models can reconcile observed line shifts and asymmetries, as well as the possible role of binary black hole systems in shaping emission-line profiles. {  The third one provided a physical framework for the interpretation of the prominent red-ward asymmetries based on gravitational redshift.} {  While the three earlier studies provided the key empirical and interpretive foundations, the present work goes further by integrating their results into a unified picture. By combining statistical trends of broad-line asymmetries with the detailed modeling of 3C 47, we show that accretion disk emission, complemented by a failed wind contribution, can consistently explain both low- and high-ionization line profiles in radio-loud quasars. Considering the similarity between Population B RQ and RL sources,  {these} results can be extended to the entire Population B. The new insight of this paper lies in connecting case-specific modeling and physical interpretation of a rare, apparently peculiar, double-peaked source with a broader view of the BLR: a predominantly virialized system whose inner edge extends to only a few hundred gravitational radii from the central black hole, making gravitational redshift the main driver of the redward asymmetry observed in the Balmer line profiles of very massive type-1 AGN.}  We further conjecture  that double peakers might have their accretion disk ``trimmed'' by a second black hole  \citep{yangetal19,mahapatraetal22}, a circumstance that might be more likely in highly evolved systems. This has significant implications for understanding the phenomenology and interpretation of quasars belonging to Population B, particularly extreme Population B, which are interpreted as mature AGN nearing a starvation phase.

We conclude that the two striking features revealed in the emission profiles of RL AGN, namely the red-ward asymmetry at the base of the line profile and the almost unshifted and symmetric \civ\ profiles, are more closely associated with large black hole mass and low radiative output than radio power and the presence of a relativistic jet.

\section*{Acknowledgments}
 
 This research has made use of the NASA/IPAC Extragalactic Database (NED),
which is operated by the Jet Propulsion Laboratory, California Institute of Technology,
under contract with the National Aeronautics and Space Administration. P. M. expresses gratitude to the organizing committee of the VIth Conference on AGN and Gravitational Lenses for the kind invitation to deliver a talk. Authors AdO, MP, JP, ADM and IM acknowledge financial support from the Spanish MCIU through project PID2022-140871NB-C21 by
“ERDF A way of making Europe”, and the Severo Ochoa grant CEX2021- 515001131-S funded by
MCIN/AEI/10.13039/501100011033.
This work is supported by the Ministry of Science, Technological Development and 
Innovation of the Republic of Serbia, contract No. 451-03-66/2024-03/200002 (N. B., E. B. and L. C\v. P.), {\textcolor{red}{and the Ethiopian Ministry of Innovation and Technology}}.
 
\bibliographystyle{jasr-model5-names}
\biboptions{authoryear}


\end{document}